\documentclass[amsmath,10t,twocolumn,superscriptaddress,footinbib,prx]{revtex4-2}
\pdfoutput=1
\usepackage{amsmath,amssymb,amsbsy,amsfonts,amsthm,bbm,bm}
\usepackage{color}
\usepackage{physics}
\usepackage{empheq}
\usepackage[colorlinks=true,allcolors=blue]{hyperref}
\usepackage[normalem]{ulem}
\usepackage{tikz}
\usepackage{xcolor}
\usetikzlibrary{calc,arrows.meta,automata,chains,positioning,shapes,matrix}
\usetikzlibrary{decorations.markings}
\usetikzlibrary{arrows, decorations.markings}
\usetikzlibrary{shapes.geometric}
\usepackage[version=4]{mhchem}
\usepackage[caption=false]{subfig}
\usepackage{mathtools}
\usepackage{yfonts}
\usepackage{stackengine}
\usepackage{mathrsfs}
\usepackage{tikz}

\definecolor{S_Blue}{RGB}{0,135,252}
\definecolor{S_Red}{RGB}{214,13,63}
\definecolor{S_Grey}{RGB}{180,180,180}

\newcommand{\bfn}{\mathbf{n}}

\usepackage{comment}

\begin{document}

\title{
Thermodynamics of large-scale chemical reaction networks
}

\author{Schuyler Nicholson}
\email{schuyler.nicholson@northwestern.edu}
\affiliation{Department of Chemistry, Northwestern University, 2145 Sheridan Road, Evanston, Illinois 60208, USA}

\author{Luis Pedro Garc\'ia-Pintos}
\affiliation{
Theoretical Division, Los Alamos National Laboratory, Los Alamos, New Mexico 87545, USA}
\date{\today}

\begin{abstract}

Chemical and biological networks can describe a wide variety of processes, from gene regulatory networks to biochemical oscillations. Modeled by chemical master equations, these processes are inherently stochastic, as fluctuations dominate deterministic order at mesoscopic scales. These classic many-body processes suffer from the so-called curse of high dimensionality, which makes exact mathematical descriptions exponentially expensive to compute. The exponential cost renders the study of the thermodynamic properties of such systems out of equilibrium intractable and forces approximations of system noise or assumptions of continuous particle numbers.
Here, we use tensor networks to numerically explore the thermodynamics of chemical processes by directly solving the ensemble solution of the chemical master equation with efficient (sub-exponential) computational cost. We provide accurate estimates of the entropy production rate, heat flux, chemical work, and nonequilibrium thermodynamic potentials, free from sampling errors or mean-field approximations. We illustrate our results through a dissipative self-assembly model. In this way, we show how tensor networks can inform the design of efficient chemical processes in previously unattainable regimes. 
\end{abstract}

\maketitle

Living processes operate far from equilibrium by expending energy and processing information at the cost of thermodynamic dissipation. At the molecular level, chemical reaction networks that transform out-of-equilibrium ‘fuel’ to ‘waste’ are the engines that power the biomolecular machinery of the cell~\cite{borsley2022chemical}. Despite residing in highly noisy environments, biological systems can precisely regulate the processing of energy~\cite{cao2015free},  timekeeping~\cite{barato2016cost,andrieux2008fluctuation,prech2025optimal}, and the expression of proteins~\cite{fange2006noise,jia2023coupling,majka2024stable}. Yet, our understanding and prediction of biological functioning remains incomplete, with simple biochemical networks leading to unforeseen dynamics~\cite{lopatkin2020predictive}. A key insight is that to understand nonequilibrium biochemical networks, one must consider not only structural complexity but also the thermodynamics of the process~\cite{dou2019thermodynamic,kriebisch2024template}. Characterizing the thermodynamics of such out-of-equilibrium processes is not only of fundamental interest but also of technological relevance. For instance, engineering efficient biomolecular machines~\cite{balzani2000artificial} or light-harvesting complexes~\cite{noy2006design, savolainen2008controlling}, or understanding some of nature's surprisingly efficient processes~\cite{howard1996structural, scholes2011lessons} rely on our ability to characterize the thermodynamics of highly complex dynamical systems.

Statistical mechanics serves as a fundamental framework for thermodynamics, by linking the behavior of atoms and molecules at the microscopic scale to the macroscopic notions of heat, work, and entropy~\cite{pathria2017statistical}. The field of stochastic thermodynamics~\cite{van2013stochastic,gaspard2022statistical,peliti2021stochastic} provides a modern treatment of nonequilibrium thermodynamics in which one can study dissipation and thermodynamic efficiency~\cite{seifert2025stochastic} at mesoscopic scales where incorporating fluctuations is integral to understanding system dynamics. However, as the number of particles in a many-body system increases, the cost of solving exactly grows exponentially. This simulation complexity is often referred to as the ``curse of dimensionality," and means that computational solutions for large biochemical networks quickly become intractable without the aid of additional approximations~\cite{cao2006efficient,gillespie2013perspective}. As a result, most approaches aim to bound thermodynamic quantities such as dissipation~\cite{horowitz2020thermodynamic,yadav2024mismatch,dechant2021improving,vo2020unified}, rather than directly calculating them.

This article's main aim is to study the thermodynamics of large-scale discrete-state chemical reaction networks, i.e., those described by the chemical master equation~\cite{gardiner2009stochastic}. To do so, we leverage tensor network techniques to directly solve the chemical master equation yielding a description of chemical reaction networks at the ensemble level. Instead of bounding thermodynamic quantities, we approximate the true thermodynamic observables using controllable tensor network approximations and efficiently computable information-theoretic quantities.

Tensor networks have been successfully applied as general function approximators across the physical sciences. From quantum computing and algorithms~\cite{chen2023quantum,tindall2024efficient,stoudenmire2024opening,berezutskii2025tensor}, to parameter estimation~\cite{liao2015tensor,ion2021tensor}, classic stochastic processes~\cite{garrahan2016classical,merbis2025effective,dolgov2024tensor}, rare events and large deviation theory~\cite{helms2020dynamical,causer2021optimal,nicholson2023quantifying} and chemical reaction networks~\cite{kazeev2015tensor,hegland2010numerical,dinh2020adaptive,kazeev2014direct,munsky2006finite,zima2025chemical}. While previous work focused on the kinetics of the chemical master equation, we will focus on the thermodynamics. To do so, we use tensor networks to probe the thermodynamic behavior of chemical reaction networks. With such an approach, we calculate heat flows, chemical work, entropy changes, and entropy production in chemical reactions as we let the volume and particle number grow towards the thermodynamic limit~\cite{higham2008modeling}, thereby gaining access to observables in previously unattainable regimes.

The manuscript is structured as follows: We review the master equations that model chemical reactions in Sec.~\ref{sec:CME} as well as the model system we use to illustrate our results. Section~\ref{sec:TN} includes a brief introduction to tensor networks and how to encode chemical reaction networks with them. Section~\ref{sec:thermodynamics} consists of an overview of stochastic thermodynamics of chemical reaction networks. Section~\ref{sec:main} contains the main results of this work, where we study the dissipation and thermodynamic efficiency of a dissipative self-assembly model as the volume and particle numbers grow.

\section{Chemical Kinetics}
\label{sec:CME}
For chemical reaction networks, the chemical master equation provides a stochastic formalism that describes the joint probability of observing each possible configuration of molecules $\bfn = [n^{X_1},n^{X_2},\ldots]$ in time~\cite{qian2010chemical}, where $n^{X_j}$ are the populations of species $X_j$.  A general chemical reaction network (CRN) for the set of reactions $r \in [\pm 1,\pm 2,\ldots, \pm N_r]$ can be written as,
\begin{equation}
\sum_{\alpha=1}^{N_\alpha}q_\alpha^{r+} n^{X_\alpha} + \sum_{j=1}^{N_j} \eta_j^{r+} n^{X_j} \rightleftharpoons
\sum_{\alpha = 1}^{N_\alpha}q_{\alpha}^{r-} n^{X_\alpha} + \sum_{j=1}^{N_j} \eta_j^{r-} n^{X_j},
\label{eq:CRN}
\end{equation}
where $n_\alpha \in \mathcal{C}$ are numbers of chemostated species held at fixed concentrations and belonging to the set $\mathcal{C}$. $n_j \in \mathcal{D}$ belong to the set of dynamical species $\mathcal{D}$. 
The $\nu_j^r = \eta_j^{r-} - \eta_j^{r+}$s and $u_\alpha^r = q_\alpha^{r-}-q_\alpha^{r+}$s are stoichiometric coefficients for reaction $r$. We use the Latin and Greek subscripts to distinguish dynamic and chemostated species.

By assuming the system is described by the chemical master equation~\cite{isaacson2006incorporating}, the joint probability $p_t(\bfn)$ of observing configuration $\bfn$ at time $t$ evolves according to the set of linear equations
\begin{equation}
\frac{dp_t(\bfn')}{dt} = \sum_{r}^N\left[\alpha_r(\bfn' - \nu^r) p_t(\bfn' - \nu^r) -\alpha_r(\bfn')p_t(\bfn')\right].
\label{eq:dpddt}
\end{equation}
Here, $\nu^r$ is the stoichiometric vector which maps state transitions $\bfn' = \bfn + \nu^r$. 
The probability per unit time that reaction $r$ occurs, known as the propensity function $\alpha(\bfn)$, is given by
\begin{equation}
\alpha_r(\bfn) = c_r\prod_\alpha n_\alpha^{q_\alpha^{r}}\prod_{X_i \in r}\frac{n_i!}{(n_i - \eta_i^{r})!},
\label{eq:Prop}
\end{equation}
such that, $\alpha_r(\bfn)dt$ is the probability that reaction $r$ will occur between $[t,t+dt)$. 
The stochastic rate coefficient $c_r$ represents the probability per unit time that reaction $r$ with state $\bfn$ will react in the next $\delta t$ time~\cite{gillespie1976general}.

Throughout this work, we adopt mass-action kinetics, which assumes that the chemostated terms of the propensity are treated differently than the dynamic terms. Terms such as $\prod_j n_j$ represent a product over all dynamical species belonging to state $\bfn$. Likewise,  $\prod_\alpha n_\alpha$ is a product over all chemostated species in state $\bfn$. Eq.~\eqref{eq:dpddt} gives the change in probability for each state $\bfn$. Equivalently we can write the change in probability as the transition matrix $H$ acting on $p_t(\bfn)$,
\begin{equation}
\frac{dp_t(\bfn')}{dt} = \sum_{\bfn,r} H^r_{\bfn'\bfn}p_t(\bfn),
\label{eq:dpH}
\end{equation}
where the propensities are now the elements of $H$, $H_{\bfn'\bfn}^r =  \alpha_r(\bfn)\delta(\boldsymbol{m} - \nu^r,\bfn)$.

\subsection*{Model system: Dissipative Self-Assembly}

To illustrate our results, we will study the driven self-assembly system~\cite{ragazzon2018energy,penocchio2019thermodynamic}, which consists of the following five sets of reversible reactions: 
\begin{align}
\ce{ F + M &<=>[c_1^+][c_1^-] M^*}, \nonumber \\
\ce{ W + M &<=>[c_2^+][c_2^-] M^*}, \nonumber \\
\ce{ $2$M^* &<=>[c_3^+][c_3^-] A^*_2}, \nonumber \\
\ce{ A^*_2 &<=>[c_4^+][c_4^-] A_2 + 2F}, \nonumber \\
\ce{ A^*_2 &<=>[c_5^+][c_5^-] A_2 + 2W}. 
\label{eq:SA_CRN}
\end{align}
Though relatively modest in reaction count, this self-assembling system poses modeling challenges due to being nonlinear at the bimolecular step. The species M, M$^*$, and A$_2^*$ are intermediate species, with $A_2$ being the target self-assembled product of the reaction. $F$ and $W$ represent fuel and waste species, which are held fixed by external chemostats. With only access to the waste species W, the monomers tend to create the activated monomer M$^*$ and disassociate back, resulting in very little of the assembled species A$_2$. The fuel is the key to assembly, driving the creation of the self-assembled structure in quantities not attainable at equilibrium. The cost of access to these new dynamic states is the inevitable dissipation of energy during the reaction~\cite{ragazzon2018energy}. 

The reaction network occupies the set of states $\bfn = [n^\text{M},n^{\text{M}^*},n^{\text{A}^*_2},n^{\text{A}_2}]$, with probability $p_t(\bfn)$. If the maximum number of species the system can generate is capped at $\mathcal{M}$, then the state space has $\mathcal{N} = (\mathcal{M}+1)^4$ elements. The amount of assembled species depends on the energy per molecule and the number of chemostated species, but also on the volume of the reaction voxel. 
In the simulations described below, we start with a monomer concentration of $1M$ ($M \equiv$ Molar concentration) for each volume we consider. This concentration corresponds to $n^\text{M} = \lfloor n_A V\rfloor$ molecules, where $n_A$ is Avogadro's constant and $V$ is the volume. The populations of all other dynamic species are initially zero, giving a distribution with probability one at state $\bfn = [n^\text{M},0,0,0]$. Due to the structure of this CRN, when starting from our initial distribution, no species can evolve to have more than $n^\text{M}$ molecules. We can then fix the maximum molecule cap to $\mathcal{M} = \lfloor n_A V\rfloor + 5$ so as not to constrain the free-evolution of the reaction. 

Note that, despite the cap on the number of molecules, the number of states the system can explore still grows exponentially. For the largest volume we consider below, $V=3e^{-22}$L, $p_t(\bfn)$ already consists of $186^4 \approx 1.29\times 10^9$ states. The Tensor network framework introduced next will allow us to study how the dissipation of the CRN relates to the external chemical work and internal thermodynamic quantities, such as heat flux and entropy, as the volume and number of particles grows.

\section{Tensor Networks}
\label{sec:TN}
Due to the indistinguishability of molecules, it is convenient to express $p_t(\bfn)$ in the Fock, or occupation basis~\cite{del2022probabilistic},
\begin{equation}
\ket{p_t} = \sum_\bfn p_t(\bfn)\ket{\bfn}.
\label{eq:pKet}
\end{equation}
Here, $\ket{\bfn} = \ket{n^{X_1}}\otimes\ket{n^{X_2}}\otimes \ldots \otimes \ket{n^{X_L}}$ is a tensor product space, in which each basis-set represents one chemical species. We use $\bra{x} = [x_1,x_2,\ldots]$ to denote the row vector of elements, and $\ket{x}$ to denote a column vector of $x$. This notation typically found in quantum physics is convenient given the mathematical framework that follows, however, we stress that all processes are classical throughout this paper.

In theory, the chemical master equation propagates a density through an infinite state space. However, in practice as illustrated by the network, Eq.~\ref{eq:SA_CRN}, one can often cap the maximum occupation number of each chemical species to $\mathcal{M}$ without significantly changing the joint distribution $p_t(\bfn)$. The ket-state $\ket{p_t}$ consists of $(\mathcal{M}+1)^L$ elements for a system of $L$ chemical species. This high-dimensional space can be decomposed into a series of low-rank tensors called a matrix product state (MPS), or tensor train~\cite{oseledets2011tensor}
\begin{align}
\ket{p_t} &\approx \nonumber \\
 &\sum_{a_1:a_{L-1}}\sum_{\bfn} C^{n_1}_{a_1}C^{n_2}_{a_1a_2}\ldots, C^{n_{L-1}}_{a_{L-2}a_{L-1}}C^{n_L}_{a_{L-1}}\ket{n_1,\ldots,n_L}\nonumber \\
 &\eqqcolon \ket{\tilde{p}_t}.
\end{align}

Each tensor $C^{n_l}_{a_{l-1}a_l}$ consists of an input, the physical index $n_l$, which represents the number of molecules of species $l$. $a_k$ are the bond-indices: fictitious indices that tie tensors together. We can visualize the MPS in a diagrammatic form as follows:
\begin{figure}[!h]
\begin{tikzpicture}[MPSStyle/.style={circle,draw,fill=S_Red,minimum size=20}]
\vspace{3mm}
\node(pwm) at (-3.1,2) {$p_t(n^{\text{X}_1},n^{\text{X}_2},\cdots,n^{\text{X}_{L}}) \approx$};
\draw[line width=.4mm] (-.5,2) -- (1.1,2);  
\node (d) at (1.5,2) {\Large $\cdots$};
\draw[line width=.4mm] (1.9,2) -- (2.5,2);  
\draw[line width=.4mm] (2.5,2) --  (2.5,1.4);
\node[MPSStyle] (se) at (2.5,2) {};
\node (s1) at (2.6,1.2) {\large $n^{\text{X}_{L}}$};
%
\foreach \x in {1,...,2}{
\draw[line width=.4mm] (-1.5+1*\x,2) -- (-1.5+1*\x,1.4);
\node[MPSStyle] (p) at (-1.5+1*\x,2) {};
\node (s1) at (-1.4+1*\x,1.2) {\large $n^{\text{X}_{\x}}$};
} 
\end{tikzpicture}
\end{figure}

\noindent The circles represent core tensors $C^{n_l}_{a_{l-1}a_l}$, with the number of lines coming from them indicating the dimension of the tensor. Connected lines are contracted over, such as the horizontal bond indices. Open indices, here, corresponding to physical indices that take the numbers of molecules as inputs, are not contracted over. 

Each element in a given bond-index represents a differing importance for re-creating the original high-rank tensor. Importance is accounted for through through the diagonal singular value-matrix $S$, whose elements are the non-negative singular values $s_{b}$, $s_1 \geq s_2 \geq \ldots $. To see this for the MPS, group the rows and columns of $C^{n_k}$ as, $C^{n_k}_{a_{k-1}a_k} = C_{(n_k,a_{k-1}),a_k}$. By contracting with a similarly ordered tensor $C^{n_{k+1}}$ and taking the singular value decomposition, we get
\begin{equation}
\sum_{a_k} C_{(n_k,a_{k-1}),a_k}C_{a_k,(a_{k+1},n_{k+1})} = \sum_{b} U^{n_k}_{a_{k-1}b}S_{bb}V^{\dagger n_{k+1}}_{ba_{k+1}}.
\end{equation}
By only keeping $D$ singular values throughout the TN we induce a total error~\cite{verstraete2006matrix,oseledets2010tt}
\begin{align}
\big\| \ket{p_t} - \ket{\tilde{p}_t} \big\|_2 \leq 2\sum_{k=1}^L \epsilon_k(D), 
\end{align}
where $\epsilon_k(D) = \sum_{b = D+1} s_b^2$ are the sum of the discarded singular values squared. 

The advantage of writing our high-rank tensor as an MPS is that one trades the $n^L$ parameters of the original tensor for $L-2$ tensors of size $nD^2$ and two at the ends of size $nD$. The total number of parameters needed to specify elements of the original tensor now grows linearly with the number of tensors $L$. Conservation of probability in the MPS format is written as the contraction of each physical index by the vectors of ones, $\ket{\boldsymbol{1}}$
\begin{figure}[h]
\begin{tikzpicture}[MPSStyle/.style={circle,draw,fill=S_Red,minimum size=20},
OneTen/.style={rectangle, rounded corners=1ex,draw,fill=S_Grey,minimum size=21}
]
\node(pwm) at (-1.6,2) {\large $\bra{\boldsymbol{1}}\ket{\tilde{p}_t} = $};
\draw[line width=.4mm] (-.5,2) -- (2.1,2);  
\node (d) at (2.5,2) {\Large $\cdots$};
\draw[line width=.4mm] (2.9,2) -- (3.5,2);  
\draw[line width=.4mm] (3.5,2) --  (3.5,1.35);
\node[MPSStyle] (se) at (3.5,2) {};
\node[OneTen,rotate=-0] (p) at (3.5,1) {$\boldsymbol{1}$};
\node (eq1) at (4.2,2) {\large $= 1$};
%
\foreach \x in {1,...,3}{
\draw[line width=.4mm] (-1.5+1*\x,2) -- (-1.5+1*\x,1.35);
\node[MPSStyle] (p) at (-1.5+1*\x,2) {};
\node[OneTen,rotate=-0] (p) at (-1.5+1*\x,1) {$\boldsymbol{1}$};
} 
\draw[line width=.4mm] (1,0) -- (1.6,0);
\node[OneTen] (ot) at (1,0) {$\boldsymbol{1}$};
\node (eq2) at (.1,0) {\large $\ket{\boldsymbol{1}} = $};
\end{tikzpicture}
\end{figure}

\noindent Having written the joint probability distribution in MPS format, we will next encode the chemical reaction network into the corresponding tensors.

\subsection*{Doi-Peliti representation}

To efficiently encode the dynamics of a CRN into a TN, we will use the Doi-Peliti formalism~\cite{doi1976stochastic,peliti1985path,rey1999asymptotic,ohkubo2013algebraic}. The Doi-Peliti formalism is most often used as a way to frame many-body problems on the way to solving path integrals for field-theoretic calculations~\cite{vastola2021solving,reyes2023general,del2022probabilistic,yang2017analytical}. Here, we will use it to model our many-body chemical system~\cite{sasai2003stochastic,rousseau2025algebraic}. Writing in a second quantized form readily allows a CRN to be represented as a TN.

Working in the occupation basis, $\ket{\bfn}$, one defines a bosonic commutation relation $[x_j,x^\dagger_k] = \delta_{jk}$, $[x_j,x_j] = [x^\dagger_j,x^\dagger_j] = 0$ and a positive semi-definite operator $\hat{n}_j = x_j^\dagger x_j$~\cite{Messiah1961Quantum}. $\hat{n}_j$ is called the number operator due to how it acts on the Fock vector,  $\hat{n}_j\ket{\bfn} = n_j\ket{\bfn}$.  Likewise the creation and annihilation operators get their names due to how they move from one Fock vector to another,
\begin{align}
\hat{x}_j^\dagger\ket{n_k} &= \ket{n_k+1}\delta_{jk},\nonumber \\
\hat{x}_j\ket{n_k} &= n_k\ket{n_k-1}\delta_{jk}, \nonumber \\
\hat{x}_j\ket{0} &= 0.
\end{align}
By using the properties of the commutator and the creation and annihilation operators, each reaction can be mapped to a transition matrix $H^{r\pm}$. For example the reaction $\ce{F + M ->[c_1^+] M^*}$ is mapped to $H^{1+} = c_1^+ n^F[x_Mx^\dagger_{M^*} - x^\dagger_Mx_M \mathcal{P}^c_{M^*}]$, where $\mathcal{P}^c_{M^*}$ is a projection operator acting on the $\ket{n_{\text{M}^*}}$ Fock space.

Due to the linearity of Eq.~\eqref{eq:dpH}, summing over all reactions gives the transition matrix for the full system dynamics $H = \sum_{r\pm} H^{r\pm}$. For further details of writing reactions into a DP form, see Refs.~\cite{nicholson2023quantifying,strand2024high}. Table~\ref{Tbl:DP} shows the Doi-Peliti representation for each reaction in the dissipative self-assembly model Eq.~\eqref{eq:SA_CRN}. For reactions that result in a net increase in the occupied Fock vector, the Hamiltonian drives transitions between occupation numbers, 
$H^{r\pm}\ket{\bfn} \rightarrow \ket{\boldsymbol{m}}$, $\boldsymbol{m} > \bfn$. Projectors are needed to enforce conservation of probability, $\bra{\boldsymbol{1}} H^{r\pm} = \bra{\boldsymbol{0}}$. The self-assembly network requires two projectors, which for the $j$th species are written as,
\begin{equation}
\mathcal{P}_j = \ket{\mathcal{M}_j}\!\bra{\mathcal{M}_j},\;\;\;\; \mathcal{\tilde{P}}_j = \mathcal{P}_j + \ket{\mathcal{M}_j-1}\!\bra{\mathcal{M}_j-1}.
\end{equation}
Subtracting each projector from the identity results in new projectors, $\mathcal{P}^c_j = \mathbb{I} - \mathcal{P}_j$, $\mathcal{\tilde{P}}^c_j = \mathbb{I} - \mathcal{\tilde{P}}_j$ that appear in Tbl.~\ref{Tbl:DP}. In the table, rate coefficients $c_1^+$, $c_2^+$, $c_4^+$ and $c_5^+$ are written as functions of the chemostated species $F,W$, $c_1^+ = \tilde{c}_1^+ n^F$, $c_2^+ = \tilde{c}_2^+ n^W$, $c_4^+ = \tilde{c}_4^+n^F(n^F-1)$, $c_5^+ = \tilde{c}_5^+n^W(n^W-1)$.

\begin{table}
\begin{tabular}{c|l}
\multicolumn{1}{c|}{Reaction} & \multicolumn{1}{c}{Doi-Peliti Form}                                         \\[2mm] \hline
\ce{F + M ->[c_1^+] M^*}     &\hspace{2mm}$\displaystyle H^{1+} = c_1^+ n^F[x_Mx^\dagger_{M^*} - x^\dagger_Mx_M \mathcal{P}^c_{M^*}]$  \\[1mm] 
\ce{M^* ->[c_1^-] F + M}     &\hspace{2mm}$\displaystyle H^{1-} = c_1^-[x_{M^*}x^\dagger_M - x^\dagger_{M^*}x_{M^*}\mathcal{P}^c_{M}]$ \\[1mm] 
\ce{M + W ->[c_2^+] M^*}     &\hspace{2mm}$\displaystyle H^{2+} = c_2^+n^W[x_Ma^\dagger_{M^*} - x^\dagger_M x_M\mathcal{P}^c_{M^*}]$     \\[1mm]
\ce{M^* ->[c_2^-] M + W}     &\hspace{2mm}$\displaystyle H^{2-} = c_2^-[x_{M^*}x^\dagger_M - x^\dagger_{M^*}x_{M^*}\mathcal{P}^c_M]$   \\[1mm] 
\ce{2M^* ->[c_3^+] A$_2$^*}  &\hspace{2mm}$\displaystyle H^{3+} = c_3^+[x^2_{M^*}x^\dagger_{A_2^*} - x^{\dagger\,2}_{M^*}x^2_{M^*}\mathcal{P}^c_{A^*_2}]$                                                        \\[1mm]
\ce{A$_2$^* ->[c_3^-] $2$M^*}&\hspace{2mm}$\displaystyle H^{3-} = c_3^-[x_{A_2^*}x^{\dagger\,2} - x_{A_2^*}^\dagger x_{A_2^*}\mathcal{\tilde{P}}^c_{M^*}]$                                                                                \\[1mm] 
\ce{$2$F + A$_2$ ->[c_4^+] A$_2$^*}&\hspace{2mm}$\displaystyle H^{4+} = c_4^+[x_{A_2}x^\dagger_{A_2^*} - x^\dagger_{A_2}x_{A_2}\mathcal{P}^c_{A_2^*}]$\\ [1mm]
\ce{A$_2$^* ->[c_4^-] A$_2$ + $2$F}&\hspace{2mm}$\displaystyle H^{4-} = c_4^-[x_{A_2^*}x^\dagger_{A_2} - x^\dagger_{A_2^*}x_{A_2^*}\mathcal{P}^c_{A_2}]$ \\[1mm] 
\ce{$2$W + A$_2$ ->[c_5^+] A$_2$^*}&\hspace{2mm}$\displaystyle H^{5+} = c_5^+[x_{A_2}x^\dagger_{A_2^*} - x^\dagger_{A_2}x_{A_2}\mathcal{P}^c_{A_2^*}]$\\ [1mm] 
\ce{A$_2$^* ->[c_5^-] A$_2$ + $2$}W&\hspace{2mm}$\displaystyle H^{5-} = c_5^-[x_{A_2^*}x^\dagger_{A_2} - x^\dagger_{A_2^*}x_{A_2^*}\mathcal{P}^c_{A_2}]$ \\[1mm] 
\end{tabular}
\caption{\label{Tbl:DP} For each reaction, there is a corresponding Doi-Peliti representation, which can be systematically found. The resulting dynamics for the whole system, $H = \sum_{r^{\pm}} H^{r\pm}$ is readily converted into a matrix product operator.}
\end{table}

Written in a Doi-Peliti form, the transition matrix is still exponentially large. For our $(\mathcal{M}+1)^L$ state system $H$ is an $(\mathcal{M}+1)^L \times (\mathcal{M}+1)^L$ matrix, a size that quickly exceeds even what can be written in an exact sparse form. By making our classical CRN resemble a quantum spin chain we have gained the ability to efficiently encode reactions as collections of low rank tensors~\cite{hubig2017generic} using pre-built packages such as autoMPO~\cite{itensor}. Due to their form, these tensor networks are called matrix product operators (MPO). In an MPO, the single high-dimensional $H$ is written in terms of $L$ low-dimensional tensors, each in the bulk consisting of two bond-indices $a_{l-1}$, $a_l$ and two physical indices, $m_l$, $n_l$. Writing out the MPO explicitly we have,
\begin{widetext}
\begin{equation}
H_{\boldsymbol{m}\bfn} \approx \sum_{\substack{m_1,\ldots,m_L \\ n_1,\ldots,n_L}}\sum_{l,a_l} W^{m_1n_1}_{a_1}W^{m_2n_2}_{a_1,a_2}\dots W^{m_{L-1}n_{L-1}}_{a_{L-2}a_{L-1}}W^{m_Ln_L}_{a_{L-1}}\ket{m_1,m_2,\ldots,m_L}\bra{n_1,n_2,\ldots,n_L}.
\label{eq:MPODef}
\end{equation}
\end{widetext}
The tensors $W^{m_ln_l}_{a_{l-1},a_l}$ constitute operator valued matrices \cite{schollwock2011density}, which here contain the creation, annihilation, and identity operators. We can again visualize the MPO Eq. \ref{eq:MPODef}, this time showing the core tensors with the blue squares,
\begin{figure}[!h]
\begin{tikzpicture}[MPOStyle/.style={rectangle,rounded corners=1ex,draw,fill=S_Blue,minimum size=20}]
\node(pwm) at (-1.55,-.45) {\large $H \approx$};
\draw[line width=.4mm] (2.9,-.5) --  (3.5,-.5);
\draw[line width=.4mm] (-.5,-.5) -- (2.1,-.5);  
\draw[line width=.4mm] (3.5,-1.15) --  (3.5,.15);
\node (d) at (2.5,-.5) {\Large $\cdots$};
\node[MPOStyle] (se) at (3.5,-.5) {};
\node (nm) at (3.6,.4) {\large $n^{\text{X}_{L}}$};
\node (mm) at (3.6,-1.4) {\large $m^{\text{X}_{L}}$};
\foreach \x in {1,...,3}{
\draw[line width=.4mm] (-1.5+1*\x,-1.15) -- (-1.5+1*\x,.15);
\node[MPOStyle] (p) at (-1.5+1*\x,-.5) {};
\node (nm) at (-1.4+1*\x,.4) {\large $n^{\text{X}_{\x}}$};
\node (mm) at (-1.4+1*\x,-1.4) {\large $m^{\text{X}_{\x}}$};
} 
\end{tikzpicture}
\end{figure}

\noindent
The bond-indices are horizontal and contracted over, while the physical indices are open. The bond-indices again bind tensors together through correlation and the physical indices represent the occupation number of each species. Thus, by feeding the states $\bfn = [n_1,n_2,n_3,\ldots,n_L]$ and $\boldsymbol{m}  = [m_1,m_2,m_3,\ldots,m_L]$ into the MPO and contracting over all tensors, one gets out the transition rate $\alpha(\bfn)$ between states $\bfn \rightarrow \boldsymbol{m}$. Just as was the case for the MPS above, the MPO does not explicitly store every transition rate, but still allows one to recall every transition rate at will.

Given an initial state $\ket{p_t}$, there are several algorithms for time evolving the state vector~\cite{paeckel2019time}. Here we use the single-site time-dependent variational principle (TDVP)~\cite{haegeman2011time,haegeman2016unifying}, as it naturally conserves probability and is capable of evolving transition matricies with longer range interactions, which are common in CRNs. With $H$ and $\ket{p_t}$ in TN form, we have the ingredients to calculate thermodynamic quantities from the chemical master equation.
\begin{figure*}[!ht]
\centering
\includegraphics[width=\textwidth]{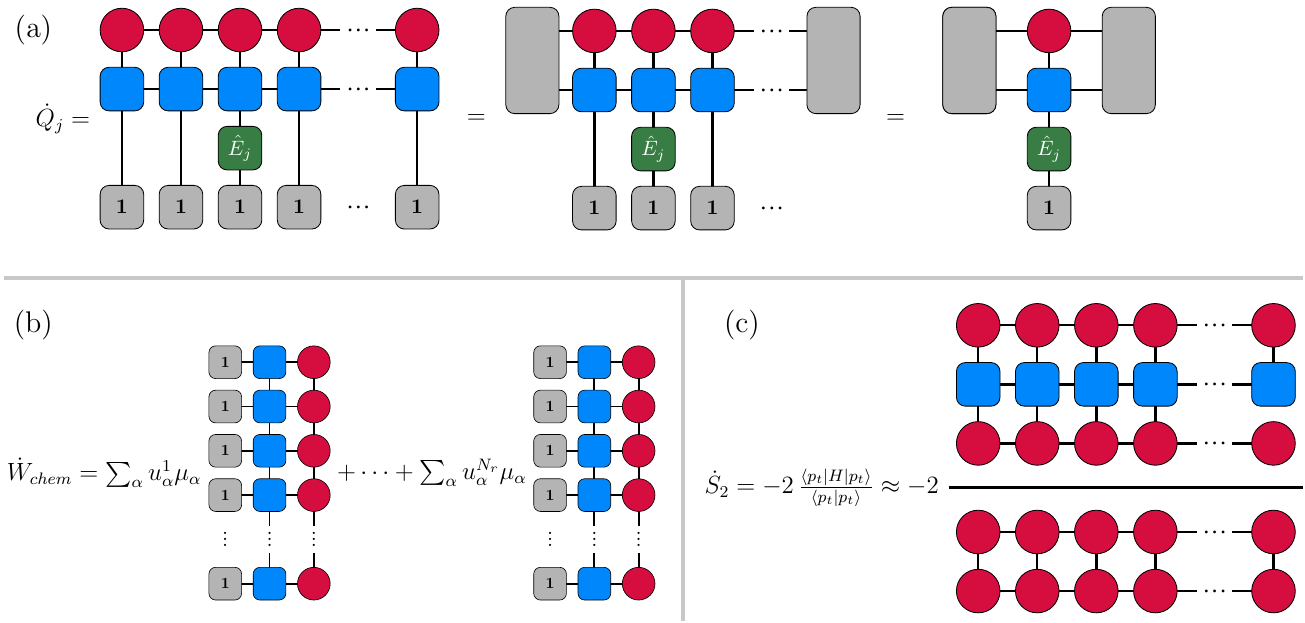}
\caption{\label{fig:TN} (a) Calculating the rate of heat flux requires the change in probability $\ket{\dot{p}_t}$. As a TN, this comes from contracting the MPO $H$ with the MPS $\ket{\tilde{p}_t}$. The energy operator only acts on one the $j$th term. In practice, it is more efficiennot toot actually calculate $\ket{\dot{p}_t}$, but to build the environment tensors (grey rectangles) from the terms $k < j$ and $k > j$, as shown by the second and third terms. 
(b) The chemical work results from the sum of MPO, MPS contractions with the independent $\ket{\boldsymbol{1}}$ tensors, each multiplied by the chemical potential and stochiometric coefficient associated with each reaction. 
(c) The Renyi-$2$ entropy is advantageous due to being the quotient of two efficient tensor network contractions. The numerator is $\bra{p_t}H\ket{p_t}$, while the denominator is the inner product $\bra{p_t}\ket{p_t}$.}
\end{figure*}

\section{Stochastic thermodynamics of chemical reactions}
\label{sec:thermodynamics}

To study the thermodynamics of a general chemical reaction network, we will focus on the central quantity of the entropy production rate (EPR) $\dot \Sigma$~\cite{seifert2025stochastic}.
The second law of thermodynamics is captured by the inequality $\dot \Sigma \geq 0$~\cite{esposito2010three}. The equality holds if a process is thermodynamically reversible, i.e., lossless. In this way, the entropy production rate quantifies the dissipation and irreversibility of a process. One measure of an efficient process is how close the integrated entropy production $\int_0^T \dot \Sigma dt$ is to zero~\cite{RevModPhys.93.035008}.

The rate of entropy production for a discrete-state continuous-time Markov process is~\cite{schmiedl2007stochastic,gaspard2004fluctuation, seifert2025stochastic}, 
\begin{equation}
\dot\Sigma = \sum_{r,\bfn',\bfn} H^{r}_{\bfn'\bfn}p_\bfn\ln\frac{H^{r}_{\bfn'\bfn}p_\bfn}{H^{r}_{\bfn\bfn'}p_{\bfn'}},
\label{eq:ms_EPR}
\end{equation}
where $H^{r} \coloneqq H^{r+} +H^{r-}$ is the sum of the rate matrices for the $r$th forward and reverse reaction. The index $r$ runs over every pair of reactions in the CRN. By summing each pair of forward and reverse reactions we enforce the local detailed balance condition Eq.~\ref{eq:ms_LDB}, that $\ln H^r_{\bfn'\bfn}/H^r_{\bfn\bfn'}$ must be zero when $\bfn' = \bfn$, see App~\ref{app:EPR} for more details. 

Equation~\eqref{eq:ms_EPR} is a purely kinetic quantity. Connections to thermodynamics are obtained when the system satisfies local detailed balance. Local detailed balance for an open CRN~\cite{rao2018conservation},
\begin{equation}
\ln \frac{H_{\bfn'\bfn}^r}{H^{r}_{\bfn\bfn'}} = -\beta\left(\Delta_r g_{\bfn'} + \sum_\alpha \mu_\alpha u_\alpha^r\right)
\label{eq:ms_LDB}
\end{equation}
is given by the log-ratio of forward and reverse rates being equal to the difference in Gibbs free-energy, $\Delta_r g_{\bfn'} = g_{\bfn'} - g_{\bfn}$, plus the chemical potential difference. The Gibbs free energy at state $\bfn$,
\begin{equation}
g_\bfn = \sum_j \epsilon_j n_j + \beta^{-1}\sum_j \ln n_j!,
\label{eq:ms_gfe}
\end{equation}
is the sum of energy per species, $\epsilon_j n_j$,  plus the internal entropic contribution $\beta^{-1}\ln n_j!$ from each species. The energy per molecule can be written $\epsilon_j = \mu_j^o - \beta^{-1}\ln n_s$, in terms of the steady state chemical potential $\mu_j^o$ and the log of the number of solvent molecules~\cite{rao2018conservation}. 

Under local detailed balance, the second law takes the form
\begin{equation}
\dot\Sigma = \dot S_{\textnormal{tot}} - \beta\dot{Q} + \beta \dot{W}_{chem} \geq 0.
\label{eq:2ndlaw}
\end{equation}
$\dot S_{\textnormal{tot}}$ is the rate of entropy change, $\dot{Q}$ is the heat flux, and $\dot{W}_{chem}$ is the rate of chemical work. 
In the next Section, we show how to write each energy contribution in a form amenable to tensor networks.

\section{Simulating the thermodynamics of large-scale chemical reactions}
\label{sec:main}

\subsection{Tensor network analysis of thermodynamic quantities}

The heat flux measures the change in energy due to changes in the distribution of states. By multiplying the energy per molecule $\epsilon_j$ by the number operator associated with species $j$, $\hat{n}_j = x_j^\dagger x_j$, one arrives at the single species energy operator, $\hat{E}_j = \epsilon_j\hat{n}_j$. When acting on a Fock state, it gives the energy contribution from species $j$ of state $\bfn$, $\hat{E}_j\ket{\bfn} = E_j\ket{\bfn}$. The heat flux can then be written compactly as,
\begin{equation}
\dot{Q} = \sum_j\bra{\boldsymbol{1}}\hat{E}_j\ket{\dot{p}}.
\label{eq:ms_qdot}
\end{equation}

The energy operator acts on a single species $j$ at a time. In general, any observable that stems from a single site operator $\hat{O}_j$ can be efficiently calculated from a TN, as illustrated in Fig.~\ref{fig:TN}a. From Eq.~\eqref{eq:dpH}, we contract the MPO and MPS together to get $d\ket{\tilde{p}_t}/dt$. The energy operator acts on site $j$, and all other entries, $k \neq j$, of the MPO are contracted with a vector of ones $\ket{1}$. To efficiently calculate a single-site observable, it's crucial to keep the dimension of intermediate tensors small. Thus, in practice, we do not construct $H\ket{\tilde{p}_t}$, but rather form the intermediate environment tensors, illustrated by gray rectangles in Fig.~\ref{fig:TN}a. For an MPS, there are left and right environment tensors, $l < j$ and $l > j$, respectively. For TNs consisting of many tensors, storing the environment tensors as one sweeps through the MPS further increases the efficiency of the calculation.

\begin{figure*}[!ht]
\centering
\includegraphics[width=1\textwidth]{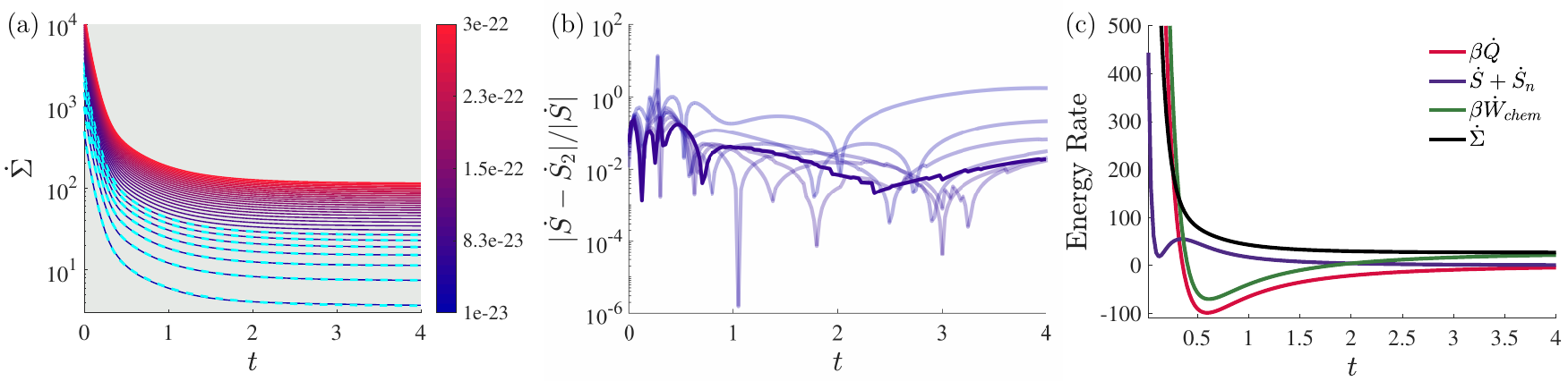}
\caption{\label{fig:cmprEPR} 
(a) Dashed blue lines represent the entropy production rate, while the solid lines approximate the entropy production rate using $\dot{\Sigma}_2$ obtained from the Renyi-$2$ entropy. 
Colors correspond to volume in liters, with the largest volume being thirty times larger than the smallest, which corresponds to a state space $\approx 4e^5$ times larger. Differences in the dissipation measures stems from using the Renyi $2$-entropy rate in place $\dot{S}$. The difference between $\dot{S}$ and $\dot{S}_2$ is large for small systems (b), but decreases as the system size grows.
(c) shows the decomposition of $\dot{\Sigma}$ into different energy rates for a volume $V = 6e^{-23}$L. Summing each contribution to the dissipation according to Eq. \eqref{eq:2ndlaw} recovers the exact dissipation rate shown by the solid black line.}
\end{figure*}

Due to every term in $H^r$ being multiplied by $u_\alpha^r$, the chemical work
\begin{equation}
\dot{W}_{chem} = -\sum_{r,\alpha}\sum_{\bfn'\bfn} H^r_{\bfn'\bfn} p_\bfn \mu_\alpha u_\alpha^r
\end{equation}
differs from the heat flux  in that $\ket{\dot{p}}$ is not directly formed. Such a difference can be attributed to $\dot{W}_{chem}$ representing the chemical work from the environment and not from an internal change in the system, like $\dot{Q}$. By defining the stoichiometric operator $\hat{u}^r$,
 \begin{equation}
\hat{u}_\alpha^r = 
\begin{cases} & u_\alpha^r\;\text{when } n'_\alpha \neq n_\alpha \\
 &0\; \text{ else}.
 \end{cases}
 \end{equation}
 we can write $\dot{W}_{chem}$ in a second-quantized notation,
\begin{align}
\dot{W}_{chem} &=  -\sum_{\alpha,r} \mu_\alpha \bra{\mathbf{1}}\hat{u}_\alpha^r H^{r}\ket{p}, \nonumber \\
&= -\sum_{\alpha,r} u_\alpha^r\mu_\alpha\bra{\mathbf{1}}\tilde{H}^{r}\ket{p}.
\end{align}
The second line is more practical for calculations since $\tilde{H}^r \coloneqq H^r - {diag}(H_{\bfn\bfn})$, the rate matrix with a zero diagonal, can be pre-calculated and easily stored as an MPO. Figure~\ref{fig:TN}c shows the TN calculation of $\dot{W}_{chem}$, where instead of an operator such as $\hat{E}_j$ acting on one site, we fully contract over the MPO representing each transition matrix $\tilde{H}^r$ with the state $\ket{\tilde{p}_t}$ and the one tensors.

For a chemical reaction network, the entropy rate is composed of two additive terms, the self-information $\dot{S} = -\bra{\ln p}\ket{\dot{p}}$ and the combinatorial uncertainty (also called the internal entropy rate) of each molecule $\dot{S}_n = -\sum_j \bra{\dot{p}}\ket{\ln n_j!}$~\cite{falasco2025macroscopic},
 \begin{align}
 \dot S_{\textnormal{tot}} = \dot{S} + \dot{S}_\bfn. 
\end{align}
Focusing on the internal entropy rate, $\dot{S}_\bfn$ can be calculated using one-site operators just as we did for $\dot{Q}$. Only now, the operator comes from forming $\ln \hat{n}_j!$ using the number operator. For systems with large physical dimensions, such as the CRN considered here, Stirling's approximation of $n_j!$ is used for large values of $n_j$. 

While $\dot{S}_\bfn$ is efficient and straightforward to calculate, the self-information $\dot{S}$ is not. The challenge comes from the need to take the logarithm of each entry of the probability distribution. Not only are there exponentially many logarithms to take, but each value of $p_t(\bfn)$ is spread over all tensors in the TN. This non-local form means we cannot easily apply local operators to take the logarithm. The result is that the cost to calculate $\dot{S}$ grows exponentially with system size, negating the efficiency gains from estimating the heat and chemical work rates. 

Instead of directly estimating the rate $\dot{S}$, we will explore quantities that can approximate it. 
The Renyi $\alpha$-entropy, 
\begin{equation}
S_\alpha = \frac{1}{1-\alpha}\ln \left(\sum_\bfn p_\bfn^\alpha\right), 
\label{eq:RenEntalpha}
\end{equation}
provides a generalization of the Shannon entropy~\cite{renyi1961measures}. It recovers the Shannon entropy $S_\alpha \rightarrow S$ for $\alpha \rightarrow 1$. For $\alpha > 1$ ($\alpha <1 $), the Renyi entropy lower (upper) bounds the Shannon entropy ~\cite{cover1999elements}. Here, we will focus on the Renyi rate,
\begin{equation}
\dot{S}_\alpha = \frac{\alpha}{1-\alpha}\frac{\bra{p_\bfn^{\alpha-1}}\ket{\dot{p}_\bfn}}{\bra{\boldsymbol{1}}\ket{p^\alpha_\bfn}}.
\label{eq:RenyiRate}
\end{equation}
For $\alpha = 2$, both $S_\alpha$ \cite{giudice2021renyi} and $\dot{S}_\alpha$ are efficiently computable using tensor networks, i.e. the computational cost grows sub-exponentially. The calculation is made up of two independent sets of TN contractions Fig. \ref{fig:TN}c.
Using $\dot{S}_2$ in place of $\dot{S}$ allows us to define a proxy dissipation rate, 
 \begin{equation}
 \dot\Sigma_2 \coloneqq \dot{S}_2 +\dot{S}_\bfn - \beta\dot{Q} + \dot{W}_{chem}.
 \label{eq:Sigdot2}
 \end{equation}
Figure~\ref{fig:cmprEPR}a compares the entropy production rate $\dot\Sigma$ to the Renyi-$2$ entropy production rate $\dot\Sigma_2$ for the self-assembly model. We start with an initial distribution that is a delta function centered at $\bfn = [n^\text{M},0,0,0]$ and evolve using TDVP up to $t = 4$. The bond dimension is fixed at $D = 30$ for volumes up to $V=1e-22$L. Above this volume, the state vector $\ket{\tilde{p}_t}$ begins to leak probability, no longer satisfying $\bra{\boldsymbol{1}}\ket{\tilde{p}_t} = 1$, meaning more singular values are required to capture the exponentially growing system size. See App. \ref{App:CmpBD} for further details. The key is that only a modest increase up to $D =60$ relative to the maximum possible bond-dimension of $(\mathcal{M}+1)^{\lfloor L/2\rfloor}$, is required to recover conservation of probability. Dashed blue lines are the exact $\dot\Sigma$ using $\dot{S}$, while solid lines are $\dot\Sigma_2$. Color denotes system size ranging from $1e^{-23}$ L to $3e^{-22}$L. For all system sizes, both dissipation measures are similar, and the approximation improves as the system size increases. The main driver of convergence is that the entropy rates $\dot{S}$ and $\dot{S}_2$ become similar with increasing system size, as shown in Fig.~\ref{fig:cmprEPR}b. 

 \begin{figure}[!h]
\centering
\includegraphics[width=.45\textwidth]{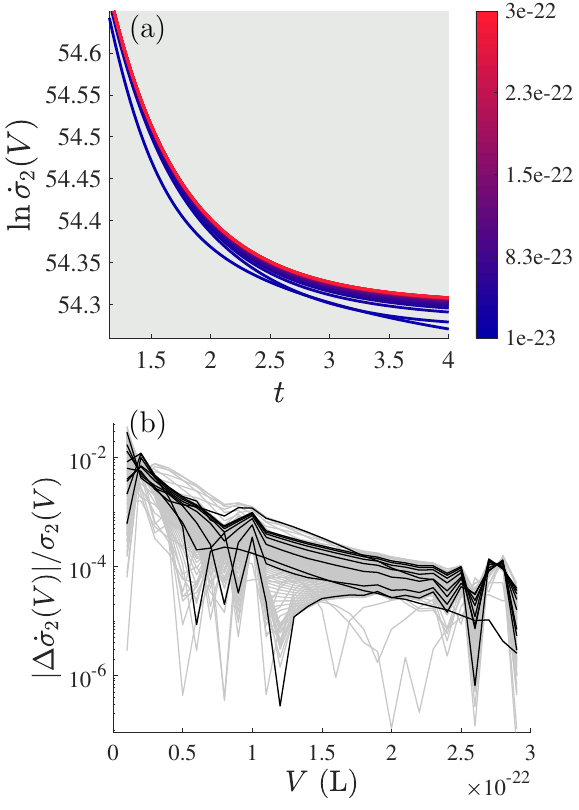}
\caption{\label{fig:Sdot_Density} (a) The EPR density for thirty volumes between $V=1e^{-23}$L and $V=3e^{-22}$L. For small systems $\sigma$ behaves unpredictably, but as the volume grows, each EPR density overlaps with each other, showing that $\sigma$ is becoming intensive and making $\dot\Sigma$ extensive. (b) Shows $\sigma$ at each time step as a function of volume. Again, as the volume increases the percent error is decreasing, but the progress is not constant. Different time steps show non-monotonic errors highlighted with a handful of representative error functions in black. Unlike asymptotic extensivity arguments of traditional thermodynamics, we see how the EPR is becoming extensive over the course of both the system's evolution and size.}
\end{figure}
A central tenet of equilibrium thermodynamics is the partition of observables into those that are extensive and those that are intensive~\cite{callen85}. However, notions of extensivity and intensivity are typically well-defined only in the thermodynamic limit~\cite{falasco2025macroscopic} and under weak interactions, not necessarily for general finite-size systems. 
By scaling the $\dot\Sigma_2$ by the system volume, we obtain an EPR density $\dot\sigma_2(V) \coloneqq \dot{\Sigma}_2V^{-1}$.  Fig.~\ref{fig:Sdot_Density}a shows that such densities differ for small system sizes. However, as the volume grows, the densities become increasingly similar. $\dot\sigma_2$ becomes intensive as the system volume increases, making $\dot\Sigma_2$ extensive. To make this statement more precise, Fig. \ref{fig:Sdot_Density}b looks at the percent error $\Delta\dot\sigma_2(V)/\dot\sigma_2$, $\Delta\dot\sigma_2 = \dot\sigma_2(V+1e-22) - \dot\sigma_2(V)$ as a function of time. As the system size grows, the error decreases at all times, though interestingly, not monotonically. For different points during the reactions, the degree of intensivity can decrease, as shown by the increase in the percent error. Tensor networks then provide a more nuanced picture than elegant thermodynamic limit arguments~\cite{prigogine1998modern,falasco2025macroscopic}.

By resolving each contribution to the EPR, our TN method demonstrates that the heat flux and rate of chemical work have larger magnitudes for these system parameters and network structure than $\dot{S}$. We can see this explicitly for the largest volume where $\dot{S}$ is calculated exactly, Fig. \ref{fig:cmprEPR}c. The relatively small $\dot{S}$ further decreases the difference between dissipation measures. For calculating the exact dissipation, the vast majority of computational time is spent on calculating the entropy rate. Due to the exponentially increasing state space, it is no surprise that the time required to calculate the entropy rate grows exponentially (App.~\ref{sect:Timings}). As a result, it becomes prohibitively expensive to simulate the larger system sizes we can tackle with the Renyi-$2$ entropy, which is accompanied by sub-exponential computational cost (App.~\ref {sect:Timings}). Being able to approximate the total dissipation generated at large volumes is one use of our method. But, having access to each term on the RHS of Eq.~\eqref{eq:Sigdot2} also allows exploration of how a system uses resources to generate a given amount of dissipation. Next, we present a detailed procedure for estimating various thermodynamic efficiency measures using tensor networks.

\subsection{Thermodynamic efficiency}

Systems evolving far from equilibrium harness reserves of information and energy to perform work and build structure. How these thermodynamic reserves are processed results in different thermodynamic functions \cite{mandal2012work,ehrich2023energy}. Here, we quantify how efficiently our CRN functions as it processes the input in chemical work and uses it to both change internal species numbers and dissipate energy. 

\begin{figure*}[!ht]
\centering
\includegraphics[width=1\textwidth]{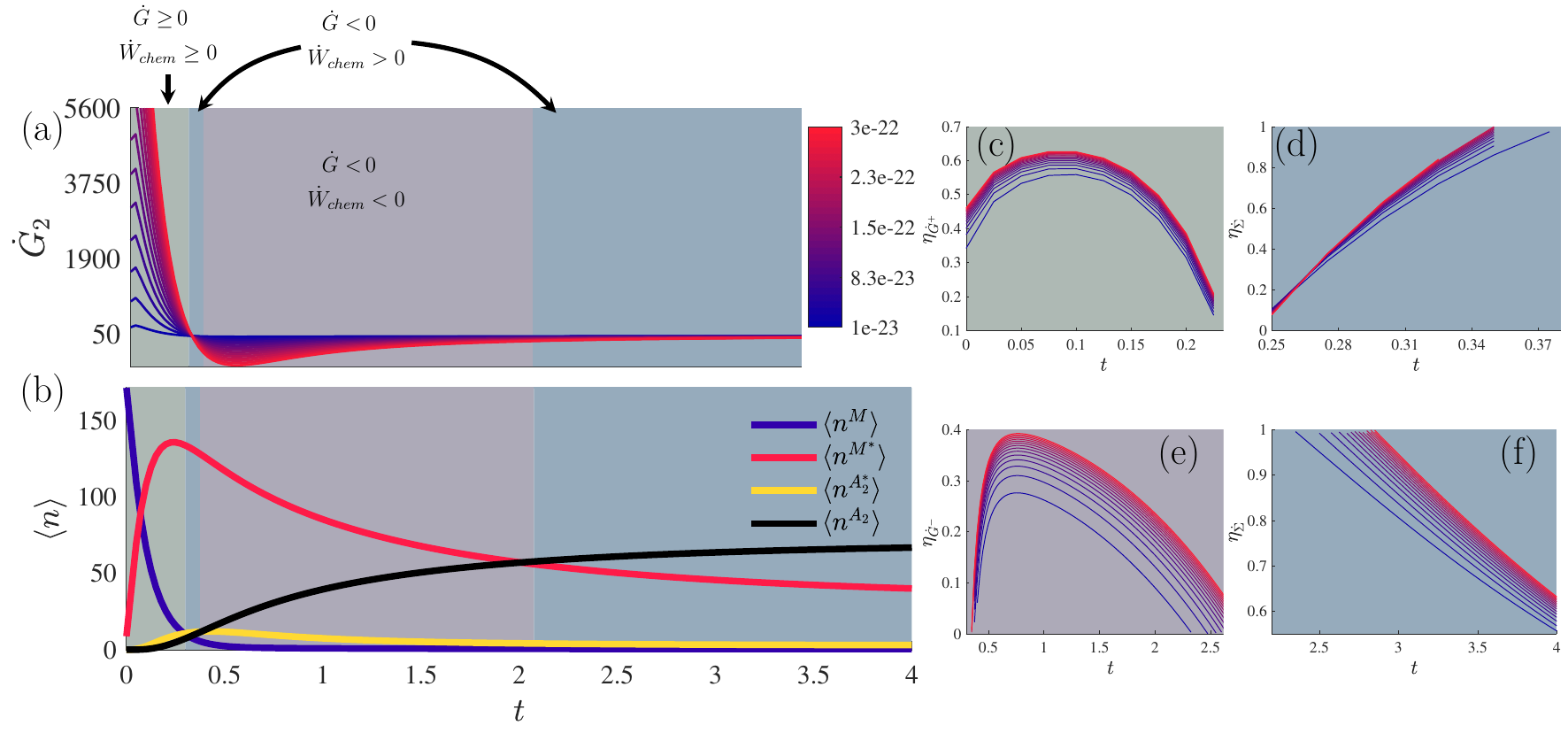}
\caption{\label{fig:Fig2} 
(a) $\dot{G}_2$ plotted for different volumes, indicated by the colorbar. As the volume increases, the magnitude of the free-energy rate increases. The background colors demarcate different thermodynamic functions undertaken by the system at $V=3e-22$L. Each period of functioning correlates to the expected number of molecules (b). Looking at the regions from left to right, when the system builds free-energy (green region), the system is growing the number of activated monomers M$^*$ with efficiency $\eta_{\dot{G}^+}$ (c).  Increasing $\eta_{\dot{G}^+}$ means less of the chemical work is being dissipated, more is being converted to free-energy. The system briefly begins using internal free-energy, blue region as $\dot{W}_{chem} > 0$, with efficiency $\eta_{\dot{\Sigma}}$ d. For the majority of time, (red region) the system is using free-energy, $\dot{G}_2 < 0$, and doing work against the environment $\dot{W}_{chem} < 0$, which corresponds to the increase in assembled structure, A$_2$. The efficiency (e) says what percent is being dissipated versus being used externally. The final phase (blue region) the, system is harnessing chemical work $\dot{W}_{chem} > 0$ and dissipating free-energy $\dot{G}_2 < 0$ in order to maintain the number of assembled molecules. An increasing efficiency (f) says that more free-energy is being dissipated with increasing volume.}
\end{figure*}

At equilibrium, we can quantify the energy that can do work as the free energy of the system~\cite{callen85}. Away from equilibrium the situation is more complicated but by grouping energy rates in the second law, one can identify a nonequilibrium analogue of the free-energy, the rate of change for the Gibbs stochastic potential $G$,
\begin{equation}
\dot{G} = \beta\dot{Q} - \dot{S} - \dot{S}_n.
\label{eq:gdot}
\end{equation}
The second law then becomes
\begin{equation}
\dot{\Sigma} = \dot{W}_{chem} - \dot{G} \geq 0.
\label{eq:2ndlawGdot}
\end{equation}
When there is no work done to modify the energies, $dE_j/dt = 0$, $\dot{G}$ is seen to stem from the rate of change in the energy representation, $G(t) = \langle E\rangle - S(t) - S_\bfn(t)$. The non-negativity of the integrated second law, $\Sigma(t) = W_{chem}(t) - G(t)\geq 0$ means that when the potential is non-negative the chemical work must also be non-negative. In this case, one can think of the system as ``charging", building a reserve of free-energy in the system~\cite{penocchio2019thermodynamic}. 

The amount of chemical work that is charged versus dissipated can be quantified through the efficiency measure
\begin{equation}
\eta_G := \frac{G}{W_{chem}} = 1 -\frac{\Sigma}{W_{chem}},\;\, 0 \leq \eta_G \leq 1.
\end{equation}
At the level of rates one can also define different efficiencies of the system. These can be divided into cases, that depend on how the system is functioning thermodynamically. 

\begin{itemize}
\item{\textit{Case 1, $\dot{G} \geq 0$:} In the first case, the system is growing the reserve of free-energy. Here, $\dot{W}_{chem}$ must also be non-negative due to $\dot\Sigma \geq 0$, and thus, $\dot{W}_{chem} \geq \dot{G}$. The efficiency $\eta_{\dot{G}^+}$ measures the amount of $\dot{W}_{chem}$ being converted to free energy versus dissipated,
\begin{equation}
\eta_{\dot{G}^+} \coloneqq \frac{\dot{G}}{\dot{W}_{chem}} = 1 - \frac{\dot\Sigma}{\dot{W}_{chem}}.
\label{eq:dotetaG}
\end{equation}
}

\item{\textit{Case 2, $\dot{G} < 0$, $\dot{W}_{chem} >  0$:} Here, the system is using its reserve of free-energy to build structure or change internal concentrations, while the environment continues to do work on the system. The sign of $\dot{G}$ now means
\begin{equation}
\dot{G} = \dot{W}_{chem} - \dot\Sigma \leq 0,
\end{equation}
which implies, $\dot\Sigma \geq \dot{W}_{chem}$. The corresponding efficiency measure describes how much free-energy is being dissipated, versus how much chemical work rate is being dissipated,
\begin{equation}
\eta_{\dot\Sigma} \coloneqq \frac{-\dot{G}}{\dot\Sigma} = 1 - \frac{\dot{W}_{chem}}{\dot\Sigma}.
\label{dotetaSigma}
\end{equation}
}

\item{\textit{Case 3, $\dot{G} <  0$, $\dot{W}_{chem} < 0$:} The third case, the system is using its reserve of free energy, but now the sign of $\dot{W}_{chem}$ says that a portion of this free-energy is doing work on the environment. Again the second law 
\begin{equation}
\dot\Sigma = \dot{W}_{chem} - \dot{G} \geq 0
\end{equation}
implies that $|\dot{G}| \geq  \dot{W}_{chem}$. The efficiency measure
\begin{equation}
\eta_{\dot{G}^-} \coloneqq \frac{\dot{W}_{chem}}{\dot{G}} = 1 + \frac{\dot\Sigma}{\dot{G}}
\end{equation}
gives the ratio of $\dot{G}$ that is being dissipated internally versus the amount being used on the environment.
}
\end{itemize}

For our self-assembling system, we will use the Renyi approximation of $\dot{G}$, which we label $\dot{G}_2 = \beta\dot{Q} - \dot{S}_2 - \dot{S}_\bfn$. We find that both free-energy measures converge as the system's size grows, allowing us to assume, $\dot{G}_2 = \dot{G}$, and to capitalize on the numerically efficiency of calculating the Renyi-$2$ entropy rate instead of the entropy rate. 

For the smallest volume we considered, $V = 1e^{-23}$, the system initially builds free-energy $\dot{G}_2 > 0$, before becoming negative and approaching zero. This is shown in Fig.~\ref{fig:Fig2}(a). With increasing volume comes increasing numbers of molecules over the system's evolution, which in turn leads to greater changes in free energy, $\dot{G} > 0$, and also longer, more pronounced period where $\dot{G} < 0$. The colored regions in Fig.~\ref{fig:Fig2} a,b show the different thermodynamic functions of the system over the course of the reaction at $V = 3e-22$L. Initially, the system converts monomers M into activated monomers M$^*$, which correspond to the green region and the building of free-energy reserves. 

Fig.~\ref{fig:Fig2}c shows that, as the system grows in size, more of the chemical work is being converted to free energy and less is being dissipated. After the number of activated monomers peaks, the system begins using its reserve of free-energy, $\dot{G}_2 < 0$, $\dot{W}_{chem} > 0$, first blue region. Fig.~\ref{fig:Fig2}d  shows that the amount of free-energy is almost all being dissipated, $\eta_{\dot\Sigma_2} \approx 1$. Thereafter,  the system shifts and, for the majority of the evolution, $\dot{G}_2 < 0$ while $\dot{W}_{chem} < 0$, which corresponds to the conversion of activated monomers into assembled molecules A$_2$ (red region).  Fig.~\ref{fig:Fig2}e gives the insight that, as the system volume grows, more of the free energy is being used externally than dissipated. The final blue region again corresponds to $\dot{G}_2 <0$ and $\dot{W}_{chem} > 0$. Here, the system uses free energy to maintain the assembled molecules A$_2$, which would otherwise revert to the expected equilibrium number without the continual dissipation of energy. (f) shows that the efficiency $\eta_{\dot\Sigma}$ is again increasing with system volume, but we also see that as time goes on, $\eta_{\dot\Sigma}$ decreases. The decrease means that physically more of the dissipation rate is coming from $\dot{W}_{chem}$ and less from $\dot{G}_2$.

The method thus allows identifying how the production of different chemical species correlates with the way the system dissipates energy and performs work in regimes otherwise inaccessible.

\section{Discussion}
\label{sec:Discussion}
Understanding how chemical and biological processes function, build structure, and respond to their environments requires understanding not just the kinetics of molecular structure and motion, but the thermodynamics of how a system processes energy and accomplishes useful work. For stochastic many-body processes, characterizing thermodynamic functioning is difficult due to the curse of dimensionality. 

The curse of dimensionality takes hold at the intermediate regime, which lies between a handful of particles and the infinite thermodynamic limit. The importance of understanding this regime lies in the many biochemical processes that operate there. Here, using tensor network methods, we show how to calculate central thermodynamic quantities such as the entropy production rate, entropy change, heat flux, chemical work, and thermodynamic efficiencies for out-of-equilibrium processes in the intermediate regime, free from sampling errors or mean-field approximations.

As the volume and number of molecules increase, a many-body system will approach the thermodynamic limit, but the rate of approach depends on the system. We have shown that we can calculate the EPR and other thermodynamic quantities in the regimes where the properties of the thermodynamic limit begin to take hold. The EPR becoming extensive with increasing system size is one such example. Being able to calculate not just the dissipation but also the ways a system dissipates energy allows us to dissect how chemical processes harvest supplies of chemical work from external reservoirs and use it to build structure. In the dissipative self-assembly model considered here, we illustrate the importance of not only measuring the dissipation produced during an evolution, but also how it relates to the formation of structure, as measured by the assembled molecules, A$_2$.

For nonequilibrium processes to generate structure and process information, there must be a dissipative price in order to satisfy the second law of thermodynamics. For open chemical reaction networks that satisfy local detailed balance, the dissipation can be decomposed into the heat flux, the rate of chemical work, and the entropy rate. But not all of these contributions are of equal magnitude. In fact, for many processes, the heat flux accounts for the majority of the dissipation~\cite{wolpert2024stochastic}. This is fortunate since each contribution the the EPR can be efficiently extracted from the MPS $\ket{\tilde{p}_t}$ except the Shannon entropy rate $\dot{S}$. Due to $\dot{S}$ being intractable to calculate for large systems, we approximated the entropy rate with the Renyi-$2$ rate. Though we show that $\dot{S}_2$ and $\dot{S}$ increasingly agree as the system size grows, a general proof of convergence is, to our knowledge, still lacking. The general framework introduced here is not limited to the self-assembly model we consider; it can be leveraged to study the thermodynamics of other chemical and biological networks away from equilibrium, crucially at sub-exponential computational cost.

\section*{Acknowledgments}

S.B.N. would like to thank Emanuele Penocchio for very helpful discussions. 
L.P.G.P.'s work was supported by the U.S. Department of Energy, Office of Science, Basic Energy Sciences program (award No. DE-SCL0000157) and the Advanced Scientific Computing Research program, under project TEQA.

\bibliography{references}


\widetext
\clearpage
\appendix 
 
 \part*{\begin{center}
\normalsize{APPENDIX  
 } 
 \end{center}}

\section{Entropy production rate}\label{app:EPR}

\noindent Here we derive the decomposition of the entropy production rate (EPR) for a chemical master equation into the individual thermodynamic rates of entropy, heat and chemical work. In the process we will find the set of equations which guarantee a collections of rates and energies are thermodynamically consistent. 

Given the dynamics is a first order continuous time Markov process, the EPR is defined as \cite{endres2017entropy,gaspard2004fluctuation},
\begin{align}
\dot\Sigma &= \sum_{r,\bfn',\bfn} H^{r}_{\bfn'\bfn}p_\bfn\ln\frac{H^{r}_{\bfn'\bfn}p_\bfn}{H^{r}_{\bfn\bfn'}p_{\bfn'}},\nonumber \\
&= -\sum_{\bfn'}\dot p_{\bfn'}\ln p_{\bfn'} + \sum_{r,\bfn',\bfn} H^{r}_{\bfn'\bfn}p_\bfn\ln\frac{H^{r}_{\bfn'\bfn}}{H^{r}_{\bfn\bfn'}}\nonumber \\
&= \dot{S} + \sum_{r,\bfn',\bfn} H^{r}_{\bfn'\bfn}p_\bfn\ln\frac{H^r_{\bfn'\bfn}}{H^{r}_{\bfn\bfn'}}.
\label{eq:Sig1}
\end{align}
The last line uses conservation of probability to recover the Shannon entropy rate. Equation~\eqref{eq:Sig1} is a purely kinetic definition of the EPR applicable to any continuous time, discrete state Markov process.
To connect to thermodynamics, we require the system to satisfy local detailed balance. Local detailed balance for an open CRN takes the form~\cite{rao2018conservation},
\begin{equation}
\ln \frac{H_{\bfn'\bfn}^r}{H^{r}_{\bfn\bfn'}} = -\beta\left(\Delta_r g_{\bfn'} + \sum_\alpha \mu_\alpha u_\alpha^r\right).
\label{eq:LDB}
\end{equation}
The Gibbs free energy of state $\bfn$ is  $g_{\bfn}$ and $\Delta_r g_{\bfn'}$ is the difference $g_{\bfn'} - g_{\bfn}$. $\mu_\alpha$ is the chemical potential of chemostated species $\alpha$ with $u_\alpha^r = q_\alpha^{r-} - q_\alpha^r$. We define
\begin{equation}
\label{eq:AuxApp1}
H^r := H^{r+} + H^{r-}
\end{equation}
where $H^{r+}$ is the rate matrix for the forward reaction and $H^{r-}$ is the rate matrix of the corresponding reverse reaction. Then the sum over $r$ in Eq.~\eqref{eq:Sig1} only runs over each pair of forward-reverse reactions. 
We can see that Eq.~\eqref{eq:AuxApp1} enforces that the LHS of Eq.~\eqref{eq:LDB} is $0$ when $\bfn = \bfn'$,
\begin{align}
 \ln \frac{H^{r+}_{\bfn\bfn} + H^{r-}_{\bfn\bfn}}{(H^{r+}_{\bfn\bfn} + H^{r-}_{\bfn\bfn})^\top} = 0   
\end{align}
When the edge, $\bfn \rightarrow \bfn'$ is non-zero for $H^{r+}_{\bfn'\bfn}$ then $H^{r-}_{\bfn'\bfn} = 0$ since by definition it represents the rate of the reverse transition. Likewise when $\bfn' \rightarrow \bfn \neq 0$ for $H^{r-}_{\bfn\bfn'}$ then $H^{r+}_{\bfn\bfn'} = 0$. The sum of $H^{r+}$ and $H^{r-}$, then guarantees that one does not divide by zero in Eq.~\ref{eq:Sig1}. Now that the kinetic form of $\dot\Sigma$ is clear, we will show how it reduces to the different rates of energy change.\\

\noindent First apply Eq.~\eqref{eq:Prop} to the LHS of Eq.~\eqref{eq:LDB} to have,
\begin{align}
&\sum_{r,\bfn,\bfn'} H_{\bfn'\bfn}^rp_\bfn\left[\ln(c_r^+\prod_\alpha n_\alpha^{q_\alpha^r}\prod_j\frac{\bfn_j!}{(\bfn_j - \eta_j^r)}) - \ln(c_r^-\prod_\alpha n_\alpha^{q_\alpha^{r-}}\prod_j\frac{\bfn'_j!}{(\bfn' - \eta_j^{r-})})\right], \nonumber \\
&= \sum_{r,\bfn,\bfn'} H_{\bfn'\bfn}^rp_\bfn\left[ \ln\frac{c_r^+}{c_r^-} -\sum_\alpha u_\alpha^r\ln n_\alpha +\sum_j\ln \frac{n_j!}{n_j'!}\right].
\label{eq:ldbLHS}
\end{align}
The second line follows from $\bfn = n + \nu^r \Rightarrow \bfn' - \eta_j^r = n + (\eta_j^{r-} - \eta_j^r) - \eta_j^r = \bfn - \eta_j^r$, canceling out the denominators in the products over $j$.\\

\noindent Next we focus on the RHS of Eq.~\eqref{eq:LDB}. The Gibbs free energy at state $\bfn$ is defined as
\begin{align}
g_\bfn &= \sum_j(\mu_j^o - \beta^{-1}\ln n_s)n_j + \beta^{-1}\sum_j\ln n_j!, \nonumber \\ 
&= \sum_j \epsilon_j n_j + \beta^{-1}\sum_j \ln n_j!
\end{align}
in terms of the energy per molecule of species $j$ ($\epsilon_j$), which depends on the standard state chemical potential $\mu_j^o$ and the number of solvent molecules in solution $n_s$~\cite{rao2018conservation}. The RHS of Eq.~\eqref{eq:LDB} can then be written as,
\begin{equation}
\ln \frac{H^{r}_{\bfn'\bfn}}{H^{r}_{\bfn\bfn'}} = -\beta\left(\sum_j \epsilon_j\nu_j^r + \beta^{-1}\sum_j \ln\frac{n_j'!}{n_j!} + \sum_\alpha \mu_\alpha u_\alpha^r\right).
\label{eq:ldbRHS}
\end{equation}
If we then equate Eq.~\eqref{eq:ldbLHS} and Eq.~\eqref{eq:ldbRHS}
\begin{equation}
\ln\frac{c^+_r}{c_r^-} = - \beta\sum_j \epsilon_j\nu_j^r + \sum_\alpha u_\alpha^r\ln n_\alpha -\beta\sum_\alpha \mu_\alpha u_\alpha^r
\label{eq:cRatio}
\end{equation}
we arrive at a representation for the ratio of forward and backward rates. Using the definition of the chemical potential for species $\alpha$, $\mu_\alpha = \mu_\alpha^o + \beta^{-1}\ln n_\alpha/n_s$, we can again identify the energy per molecule, $\epsilon_\alpha = \mu_\alpha^o - \beta^{-1}\ln n_s$, giving $\mu_\alpha = \epsilon_\alpha + \beta^{-1}\ln n_\alpha$. By plugging $\epsilon_\alpha$ into Eq.~\eqref{eq:cRatio} we have
\begin{align}
\ln\frac{c_r^+}{c_r^-} &= - \beta\sum_j \epsilon_j\nu_j^r - \beta\sum_\alpha \epsilon_\alpha u_\alpha^r, \nonumber \\
\ln \frac{c_r^-}{c_r^+} &= \beta\sum_j \epsilon_j\nu_j^r + \beta\sum_\alpha \epsilon_\alpha u_\alpha^r
\label{eq:LDBConst}
\end{align}
which recovers the set of equation that need to be satisfied to ensure the a CRN is thermodynamically consistent \cite{schmiedl2007stochastic}. Not all sets of chemical rates and energies will satisfy local detailed balance for a given CRN. Solving Eq.~\eqref{eq:LDBConst} for the set of energies, $\{\epsilon_j,\epsilon_\alpha\}$ and rates $\{c_r\}$ we ensure the system satisfies local detailed balance. Using our new relation for the ratio of rates in Eq.~\eqref{eq:ldbLHS} gives,
\begin{align}
\ln \frac{H_{\bfn'\bfn}^r}{H_{\bfn\bfn'}^{r}} &= -\beta\sum_j \epsilon_j\nu_j^r -\beta\sum_\alpha \mu_\alpha u_\alpha^r + \sum_\alpha u_\alpha^r\ln n_\alpha,\nonumber \\
&- \sum_\alpha u_\alpha^r\ln n_\alpha + \sum_j \ln \frac{n_j!}{n_j'!},\nonumber \\
&= -\beta\sum_j \epsilon_j\nu_j^r + \sum_j\frac{n_j!}{n_j'!} - \beta\sum_\alpha \mu_\alpha u_\alpha^r.
\end{align}
Finally, summing over $H^r_{\bfn'\bfn}p_\bfn$ we recover 
three thermodynamic quantities
\begin{align}
\sum_{r,\bfn',\bfn}H_{\bfn'\bfn}^r p_\bfn\ln \frac{H^r_{\bfn'\bfn}}{H^{r}_{\bfn\bfn'}} &= \sum_{r,\bfn',\bfn} H_{\bfn'\bfn}^r p_\bfn\left(-\beta\sum_j \epsilon_j\nu_j^r + \sum_j \ln \frac{n_j!}{n_j'!} - \beta\sum_\alpha \mu_\alpha u_\alpha^r\right),\nonumber \\
&= -\beta\sum_{r,\bfn',\bfn,j}H_{\bfn'\bfn}^r p_\bfn \epsilon_j(\bfn'-\bfn) - \sum_{\bfn',j}\dot p_{\bfn'}\ln n_j'! - \beta\sum_{r,\bfn',\bfn,\alpha}H_{\bfn'\bfn}^r p_\bfn \mu_\alpha u_\alpha^r.
\end{align}
 The first term is the rate of heat flux, the second is the combinatorial contribution to the entropy rate and the third is the rate of chemical work. Keeping track of the state to state transitions once we sum over $H^r_{\bfn'\bfn'}p_{\bfn}$ can be onerous. Just as we simplified the set of states $p(\bfn)$ in a second quantized form, we treat the thermodynamic quantities in the same way,\\

\noindent Writing the joint distribution with respect to the occupation basis $\ket{\bfn} = \ket{n_1,n_2,\ldots}$, $\ket{p} = \sum_\bfn p(\bfn)\ket{\bfn}$, we can define the number operator $\hat n_j$ such that $\hat n_j\ket{\bfn} = n_j\ket{\bfn}$, and the energy operator $\hat E_j$, $\hat E_j\ket{\bfn} = E_j\ket{\bfn}$. The heat flux can then be written as,
\begin{align}
 \dot{Q} &=\sum_{r,\bfn',\bfn,j}H_{\bfn'\bfn}^r p_\bfn \epsilon_j\nu_j^r, \nonumber \\
  &=  \sum_{r,j}\beta\bra{\mathbf{1}}\hat E_jH^r_{\bfn'\bfn}\ket{p}, \nonumber \\
  &= \sum_{j}\bra{\mathbf{1}}\hat E_j\ket{\dot p}.
 \end{align}
The total rate of entropy comes from combining $\dot S$ with the combinatorial entropy contribution,
\begin{equation}
\dot S_{\textnormal{tot}} = \dot S -\sum_j \bra{\ln \hat{n}_j!}\ket{\dot p}.
\label{eq:Sdot}
\end{equation}
In the rate of chemical work we will write an operator in place of the coefficients $u_\alpha^r$, where,f 
\begin{equation}
\hat{u}_\alpha^r = 
\begin{cases} & u_\alpha^r\;\text{when } n'_\alpha \neq n_\alpha \\
 &0\; \text{ else}.
 \end{cases}
 \end{equation}
Which leads to the chemical work taking the form,
\begin{align}
\dot{W}_{chem} &=  -\hspace{-2mm}\sum_{r,\bfn',\bfn,\alpha}H_{\bfn'\bfn}^r p_\bfn (\mu_\alpha u_\alpha^r), \nonumber \\
&= -\sum_{\alpha,r} \mu_\alpha \bra{\mathbf{1}}\hat{u}_\alpha^r H^{r}\ket{p}.
\label{eq:WdotChem}
\end{align}
Putting each thermodynamic component back into Eq.~\eqref{eq:Sig1} we see that the EPR can be decomposed into different rates of energy change,
\begin{equation}
\dot\Sigma = \dot S_{\textnormal{tot}} -\beta\dot Q + \beta\dot{W}_{chem}.
\label{eq:Sigdot}
\end{equation}

\section{Thermodynamically consistent reaction networks:}

For a dilute solution governed by coupled chemical reactions, the chemical master equation provides a full description of intrinsic fluctuations. To understand the thermodynamics of such as system it is critical that the dynamics be thermodynamically consistent~\cite{schmiedl2007stochastic}. Thermodynamic consistency is enforced through the principle of local detailed balance. 
Mathematically a collection of rates and energies are thermodynamically, consistent if they provide a solution to
\begin{equation}
\beta\sum_j \epsilon_j\nu_j^r = \ln\frac{c_r^-}{c_r^+}-\beta\sum_\alpha \epsilon_\alpha u_\alpha^r.
\label{eq:Consistent}
\end{equation}
for each pair of reactions $r^+$, $r^-$. $\epsilon_j$ is the energy per molecules of dynamical species $j$ and $\epsilon_\alpha$ is the energy per molecules of chemostated species $\alpha$. Eq.~\eqref{eq:Consistent} represents a set of coupled equations. For the  dissipative self-assembly network in Eq.~\eqref{eq:SA_CRN}, the set of equations takes the form 
\begin{equation}
\begin{bmatrix}
\epsilon_{M^*} - \epsilon_M  \\ \epsilon_{M^*} - \epsilon_M  \\ \epsilon_{A_2^*} - 2\epsilon_{M^*} \\ \epsilon_{A_2^*} - \epsilon_{A_2} \\ \epsilon_{A_2^*} - \epsilon_{A_2}  \end{bmatrix}
= 
\begin{bmatrix} \ln c_1^+/c_1^- \\  \ln c_2^+/c_2^- \\  \ln c_3^+/c_3^- \\  \ln c_4^+/c_4^- \\
 \ln c_5^+/c_5^-\end{bmatrix}
 -
 \begin{bmatrix} - \epsilon_F\\ -\epsilon_W \\ 0 \\ 2\epsilon_F \\ 2\epsilon_W \end{bmatrix},
 \label{eq:LDB_Mat}
 \end{equation}
 where we have assumed units such that $\beta = 1$.
 Using lines $1,2$ and $4,5$  we see that the difference in energies between the fuel and waste molecules are what drives the system away from equilibrium,
\begin{equation}
\Delta \epsilon_{WF} = \epsilon_W-\epsilon_F = \ln \frac{c_1^- c_2^+}{c_1^+c_2^-} =\frac{1}{2} \ln \frac{c_5^- c_4^+}{c_5^+ c_4^-}.
\label{eq:delEWF}
\end{equation}
If the fuel and waste energies are equal, and the system is at the equilibrium concentrations then no directed reactions take place. Being thermodynamically consistent requires that the forward and reverse rates of reactions 1 and 2, as well as 4 and 5 are also balanced. Starting from Eq.~\eqref{eq:delEWF}, we can define thermodynamically consistent energies and rates. The stochastic rate functions can be written in terms of the more experimentally accessible kinetic rate coefficients, $k_r^{\pm}$,
\begin{equation}
c^\pm_r = k^\pm_r/(n_AV)^{m-1},
\label{eq:StochRate}
\end{equation}
where $m$ are the number of reactant molecules in reaction $r$, $V$ is the system's volume and $n_A$ is Avogadro's constant. Using Eq.~\eqref{eq:StochRate} in Eq.~\eqref{eq:delEWF}, we find that the local detailed balance condition can be written in terms of the kinetic rates as,
\begin{align}
\Delta \epsilon_{WF} = \ln \frac{k_1^- k_2^+}{k_1^+k_2^-} =\frac{1}{2} \ln \frac{k_5^- k_4^+}{k_5^+ k_4^-}.    
\end{align}
Fixing ${k_1^\pm,k_2^{\pm},k_4^-}$, we can find $k_4^+$,
$$
k_4^+ = k_4^- e^{-\ln k_5^-/k_5^+ + 2\Delta E_{WF}}.
$$
Even with the detailed balance condition the energies for this model are over-determined. We choose to fix $\epsilon_W$ and $\epsilon_M$. Then, using the set of rates, $[c_1 ^\pm, c_2^\pm, \ldots, c_5^\pm]$ and Eq. \ref{eq:LDB_Mat},  we can solve for the remaining energies
\begin{align}
\epsilon_\text{F} &= \ln \frac{c_2^-}{c_2^+} + \epsilon_\text{W} + \epsilon_{\text{M}},\nonumber \\
\epsilon_{\text{M}^*} &= \ln \frac{c_2^-}{c_2^+} + \epsilon_\text{W} + \epsilon_{\text{M}}, \nonumber \\
\epsilon_{\text{A}_2^*} &= \ln\frac{c_3^-}{c_3^+} + 2\left(\epsilon_\text{F} + \epsilon_\text{M} + \ln \frac{c_1^-}{c_1^+}\right),\nonumber \\
\epsilon_{\text{A}_2} &= \ln \frac{c_5^-c_3^-}{c_5^+c_3^+} + 2\left(\ln\frac{c_2^-}{c_2^+} + \epsilon_\text{M}\right).
\end{align}
Together the set of energies and rates give a set of thermodynamically consistent parameters for the reaction network Eq.~\eqref{eq:SA_CRN}. All results were found using, $k^{r\pm} = [k_1^+,k_1^-,k_2^+,k_2^-,k_3^+,k_3^-,k_4^+,k_4^-,k_5^+,k_5^-] = [5, 3.63e^{-2}, 1e^{-3}, 6.06e^{-2}, 1, 2.27, 1e^{-8}, 1.74e^{-2}, 5, .124, .1, 4.81^{e-4}]  $ which are roughly equivalent to the rates used in \cite{penocchio2019thermodynamic} for analyzing the deterministic behavior of the system.

\section{Comparing bond dimensions}\label{App:CmpBD}

The bond dimension controls the level of approximation of a tensor network. For an MPS, if one lets the bond dimension grow to be $(\mathcal{M}+1)^{\lfloor L/2\rfloor}$, then the MPS will exactly equal the system it is encoding. 
Because the approximation made by tree-TNs is controlled, growing the bond dimension provides an increasingly accurate approximations of the underlying system. 
We can compare the TNs as bond dimension increases to get an understanding of how accurate our approximation is. 

\begin{figure}[!h]
\centering
\includegraphics[width=.75\textwidth]{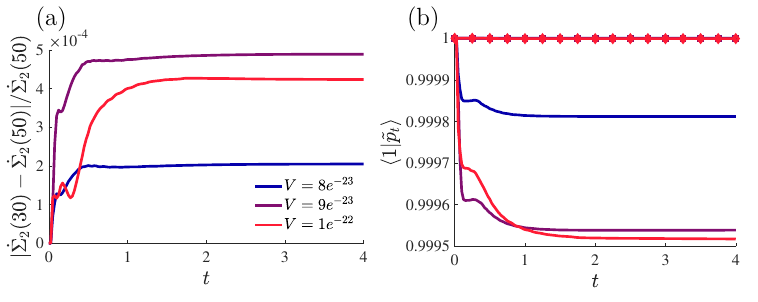}
\caption{\label{fig:CmprBD} (a) The percent difference in $\dot\Sigma_2$ using a bond dimension of $D=30$ and $D= 50$. We see that, as the volume grows, the difference in dissipation rates grows due to the larger system requiring more singular values to accurately capture the EPR. The modest percent difference at larger volume, despite using significantly less singular values, shows that the truncation is not removing significant singular values. (b) Conservation of probability is another vital measure of how well the TN is approximating the actual probability distribution. The solid lines illustrate how as the system size grows, eventually the BD must be increased to conserve probability. The lines with symbols show the total probability of the TN in time when $D=50$.}
\end{figure}

For the CRN in Eq.~\eqref{eq:SA_CRN}, we compare the dissipation estimated from a $D=30$ bond dimension TN and a $D=50$ bond dimension TN at three different volumes $V=[8e^{-23},9e^{-23},1e^{-22}]$ L. Increasing the volume increases the difference in the EPR, but not appreciably. Having a small difference in EPR for $40\%$ reduction in singular values means that the truncation at $D=30$ is not truncating singular values that are significant in measuring the EPR. Conservation of probability is our main diagnostic measure for if the probability distribution is being faithfully represented. Fig.~\ref{fig:CmprBD}b shows (solid lines) that the TN eventually starts to loose conservation of probability for large enough volume, holding the bond dimension fixed. Increasing the bond-dimension to $D=50$ recovers probability conservation as shown by the solid lines with symbols. 

\section{Runtime to numerically calculate $\dot{S}$}\label{sect:Timings}
\begin{figure}[!h]
\centering
\includegraphics[width=.45\textwidth]{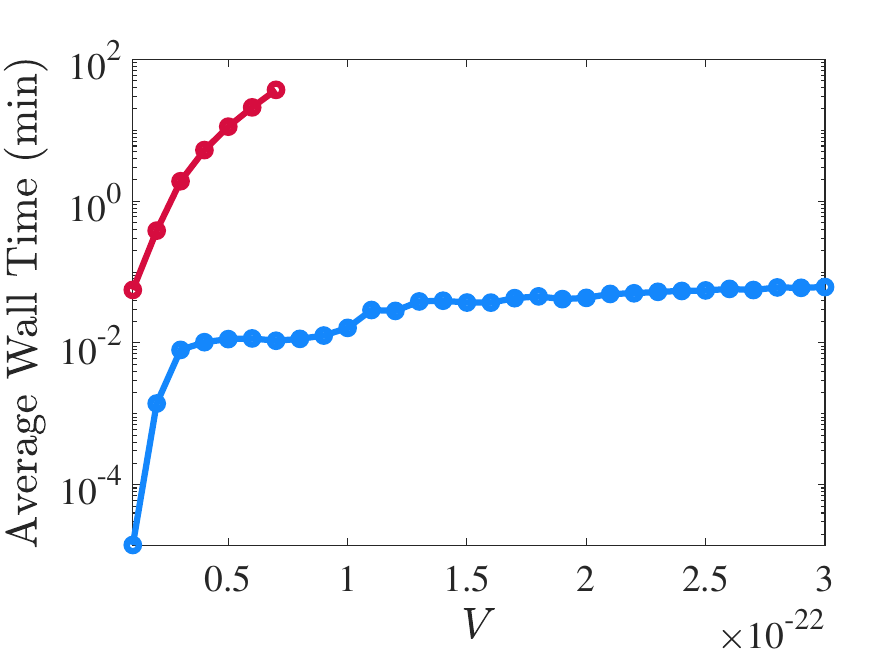}
\caption{\label{fig:ExpWallTime} Comparison of the runtime to numerically exactly simulate the entropy rate (red curve) for one time $t$, versus the runtime to estimate the entropy rate by tensor networks (blue curve) at the same time $t$. The total time for a reaction is then the estimated by the average time, multiplied by the total number of time-steps.
}
\end{figure}

We calculate the exact entropy rate by finding all values of $p_t(\bfn)$ and $\dot{p}_t(\bfn)$ from the TN at each moment in time. This is done by contracting a delta function representing each state $\bfn$ with $\ket{\tilde{p}_t}$ and $d\ket{\tilde{p}_t}/dt$. By measuring directly from the MPS, we do not include the cost of representing the large state space as a matrix or implementing a matrix exponential to time-evolve the probability distribution. The time we report to calculate $\dot{S}$ is then a kind of best case scenario. The change in probability $\dot{p}_t$ is found by contracting the MPO $H$ with the MPS $\ket{\tilde{p}_t}$. As expected, the exponentially growing state space results in an exponential increase in the average wall time to compute $\dot{S}$, as shown by the red curve in Fig.~\ref{fig:ExpWallTime}. The Renyi-$2$ entropy can be calculated through two TN calculations, Fig.~\ref{fig:TN}c, which can be efficiently computed. The result, blue curve on Fig.~\ref{fig:ExpWallTime}, is that the run time to calculate the Renyi-$2$ two-entropy grows sub-exponentially compared with to the cost of the exact change in entropy. For $V = 7e-23$L it took on average $37.16$ min per time step to calculate $\dot{S}$, but less than a second to calculate $\dot{S}_2$. The poor scaling for $\dot{S}$ leads to an unreasonable total calculation time, for the largest volume we consider with TNs, $V = 3e-22$L.

\end{document}